\documentclass[prb,preprint,showpacs,preprintnumbers,amsmath,amssymb]{revtex4}

\usepackage{graphicx}
\usepackage{dcolumn}
\usepackage{bm}

\begin{document}

\preprint{APS/123-QED}

\title{Determination of the micromagnetic parameters in (Ga,Mn)As using domain theory }

\author{C. Gourdon}
\email[e-mail: ]{catherine.gourdon@insp.jussieu.fr}

\homepage[web site: ]{http://www.insp.upmc.fr}
\author{A. Dourlat}
\author{V. Jeudy}
\author{K. Khazen}
\author{H. J. von Bardeleben}
\affiliation{
Institut des Nanosciences de Paris\\ Universit$\acute{e}$ Pierre et Marie Curie~-~Paris~6, CNRS, UMR~7588\\ 
140 rue de Lourmel, 75015 Paris, France}
\author{L. Thevenard, A. Lema\^{\i}tre}
\affiliation{Laboratoire de Photonique et Nanostructures, CNRS, UPR 20\\ Route de Nozay, 91460 Marcoussis,
France}

\date{\today}

\begin{abstract}
The magnetic domain structure and magnetic properties of a ferromagnetic (Ga,Mn)As epilayer with perpendicular magnetic easy-axis are investigated. We show that, despite strong hysteresis, domain theory at thermodynamical equilibrium can be used to determine the micromagnetic parameters. Combining magneto-optical Kerr microscopy, magnetometry and ferromagnetic resonance measurements, we obtain the characteristic parameter for magnetic domains $\lambda_c$, the domain wall width and specific energy, and the spin stiffness constant as a function of temperature. The nucleation barrier for magnetization reversal and the Walker breakdown velocity for field-driven domain wall propagation are also estimated.
\end{abstract}


\maketitle


In thin layers based on ferromagnetic (FM) metallic or semiconducting materials domain wall (DW) propagation is the subject of extensive research with the ultimate aim of information storage and/or transport by domain walls.\cite{yamaguchi,vernier,thomas,hayashi,yamanouchi2004,yamanouchi,tang,dourlat-spintech} In particular, the FM semiconductor (Ga,Mn)As has received a lot of attention recently since the critical current for DW propagation was shown to be two orders of magnitude smaller than in metallic systems.\cite{yamanouchi2004} Both current-driven and field-driven DW propagation have been recently investigated in (Ga,Mn)As layers and stripes.\cite{yamanouchi,tang,dourlat-spintech} In this context, it is important to determine experimentally the micromagnetic parameters such as the DW width and spin stiffness constant since they play a crucial role in DW dynamics. In the hydrodynamic regime for field-driven DW propagation, the DW mobility and the Walker breakdown velocity are proportional to the DW width,\cite{malozenoff-slonczewski} which depends on the spin stiffness constant and the magnetic anisotropy constants.\cite{hubert} Spin stiffness also governs the exchange length, which characterizes the Bloch line width and sets an upper limit for the size of superparamagnetic particles. Moreover, the determination of the spin stiffness constant also gives access to the $J_{pd}$ exchange parameter characterizing the $p-d$ exchange interaction between localized Mn spins and itinerant carriers (holes).\cite{dietlPRB2001,jungwirthReview,konig} The experimental determination of the spin stiffness is not straightforward. It was extracted from spin wave resonances despite the discrepancy between theoretical and experimental dispersion curves.\cite{goennenwein,konig} The obtained value was larger than the predicted ones. In this paper, we determine the spin stiffness constant and the other micromagnetic parameters using combined experimental techniques and analysing the magnetic domain structure and the hysteresis cycles in the framework of domain theory.

Owing to the complex structure of the valence band in zinc-blende (Ga,Mn)As, the direction of the magnetic easy-axis can be set parallel or perpendicular to the plane by tuning the layer strain and carrier concentration.\cite{dietlPRB2001,sawicki,thevenardAPL2005,dourlatJAP,thevenardPRB2007} In the case of strong perpendicular anisotropy, the ground state of the FM layer at thermodynamical equilibrium corresponds to the self-organized periodic stripe or bubble pattern of the up- and down-magnetized domains.\cite{hubert} Domain theory derived from the micromagnetic equations for FM films provides a quantitative description of the equilibrium domain patterns. Long-range self-organization results from the competition between the DW energy and the magnetostatic energy. The domain pattern is controlled by a single parameter $\lambda_c =\ell_c/d=\sigma/2K_d d$, where $\ell_c$ is a characteristic length equal to the ratio of the specific wall energy $\sigma$ and the magnetostatic energy with $\sigma=4\sqrt{AK_u}$ and $K_d=\mu_0 M_s^2/2 $. $A$ is the spin stiffness constant, $K_u$ the uniaxial anisotropy constant, $M_s$ the saturation magnetization, and $d$ the sample thickness. Experimentally, the parameter $\lambda_c$ can be extracted from the zero-field period of domain array. This method was used by T. Dietl \textit{et al.} to determine $\lambda_c$ for perpendicular-axis (Ga,Mn)As layers showing magnetic domains.\cite{dietl-konig-macdonald} However, the question arises whether the observed domain structures correspond to the equilibrium state. Actually, for perpendicular-axis (Ga,Mn)As the hysteresis cycle is square,\cite{thevenard,dourlatJAP} thereby indicating that metastable states and/or DW pinning play an important role in the field dependence of the magnetization. Defect-assisted nucleation processes and DW pinning were recently investigated using magneto-optical imaging.\cite{thevenard,dourlatJAP,wang2007} In particular it was reported that the equilibrium demagnetized stripe or bubble state at zero applied field cannot be reached using AC demagnetization.

In this paper we show that domain theory can be applied to determine upper and lower boundaries for the parameter $\lambda_c$ even in the presence of metastability and DW pinning. The upper boundary is obtained by comparing the domain width close to magnetization saturation with the predicted equilibrium width. The lower boundary is obtained from the comparison of major and minor hysteresis loops with the equilibrium magnetization curve.
Using the saturation magnetization and the anisotropy constants obtained by superconducting quantum interference device (SQUID) magnetometry and ferromagnetic resonance (FMR) experiments, respectively, we obtain the temperature dependence of the lower and upper boundaries for the micromagnetic parameters: the specific DW energy, the DW width, and the spin stiffness constant, which are determined within a narrow range.
 
The sample was prepared by molecular beam epitaxy.
It consists of a 50 nm thick Ga$_{0.93}$Mn$_{0.07}$As layer grown on a Ga$_{0.902}$In$_{0.098}$As relaxed buffer layer deposited on
a semi-insulating (001) GaAs substrate. After post-growth annealing the Curie temperature $T_C$ was 130 K. The magnetic domain structure was investigated using magneto-optical Kerr (MOKE) microscopy. The hysteresis cycle was obtained from the average intensity of MOKE images as a function of the applied field. Experimental details can be found elsewhere~\cite{dourlatJAP}.

\begin{figure}
	\begin{center}
		\includegraphics[width=0.95\linewidth]{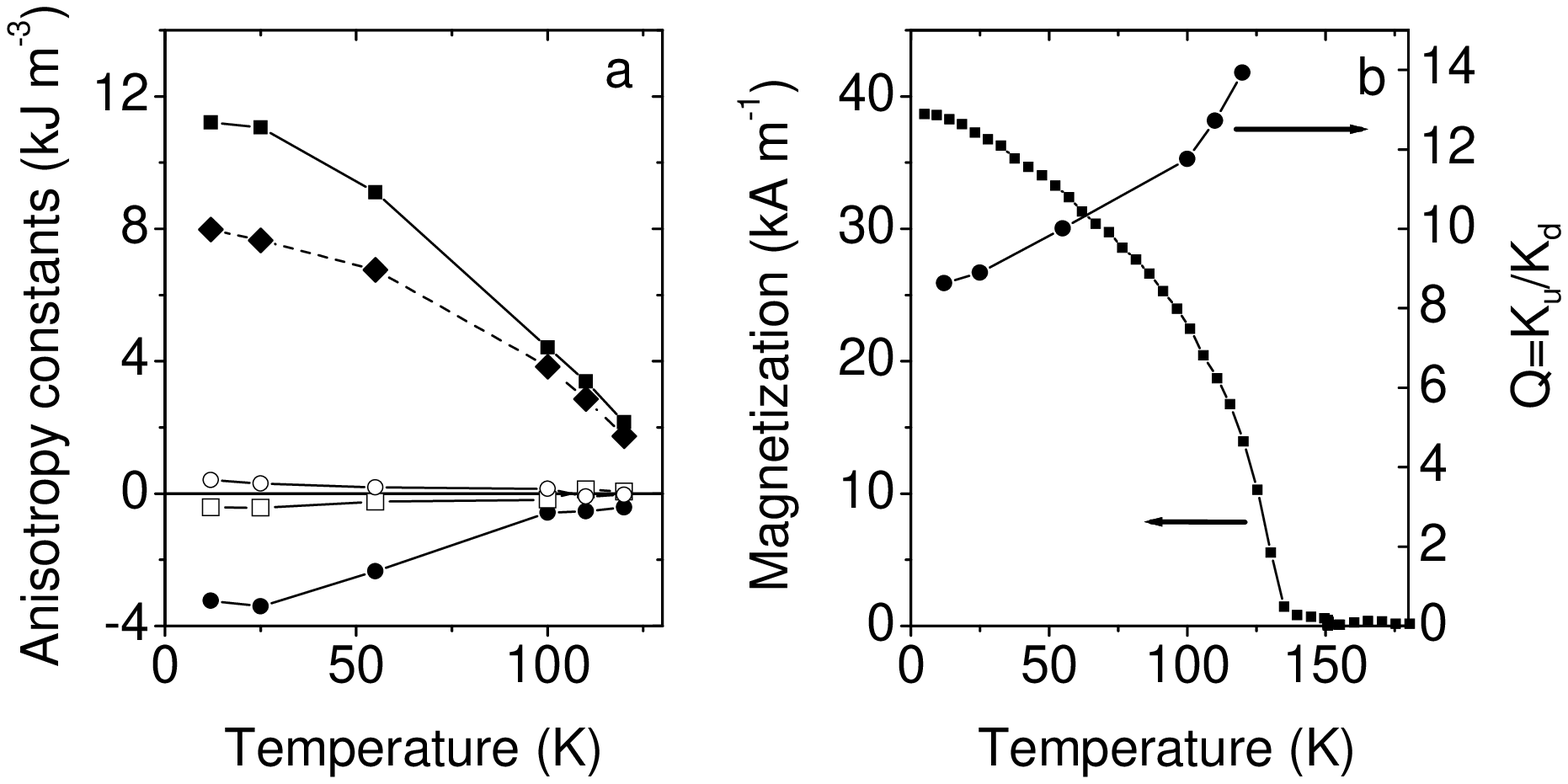}
		\end{center}
	\label{fig:Graph-K-Q}
\end{figure}

The anisotropy constants were obtained from FMR measurements. The FMR spectra were measured with X-band and Q-band spectrometers in the first derivative absorption mode with a modulation frequency of 100~kHz. To determine the anisotropy constants the spectra were measured for both the in-plane and out-of-plane orientation of the static magnetic field. From the resonance field positions we can deduce that the easy-axis of the sample is the along [001] direction for all temperatures $T<T_C$. For the in-plane angular variation of the magnetic field an almost isotropic behavior is observed. Using the Smit-Beljers equation and the minimization of the free energy for different alignments of the magnetization~\cite{liu,smit} the magnetic anisotropy constants  $K_{2\bot}$, $K_{4\bot}$, $K_{2\|}$, and $K_{4\|}$ were obtained. They are plotted in Fig.~\ref{fig:Graph-K-Q}(a) as a function of temperature. The small values of the in-plane anisotropy constants $K_{2\|}$ and $K_{4\|}$ explain the almost isotropic in-plane behavior. In the following we will neglect them and consider the first term of the development of the anisotropy energy $K_u sin^2\theta$ with $\theta$ the angle between the [001] direction and the magnetization. We obtain $K_u$ from the FMR results as $K_u=K_{2\bot}+K_{4\bot}$. The temperature dependence of $K_u$ is plotted in Fig.~\ref{fig:Graph-K-Q}(a). The uniaxial anisotropy is characterized by the parameter $Q=K_u/K_d$. From FMR and SQUID results we obtain a Q parameter ranging from 8.6 to 14 (Fig.~\ref{fig:Graph-K-Q}(b)).



\begin{figure}[htbp]
\includegraphics[width=0.5\textwidth]{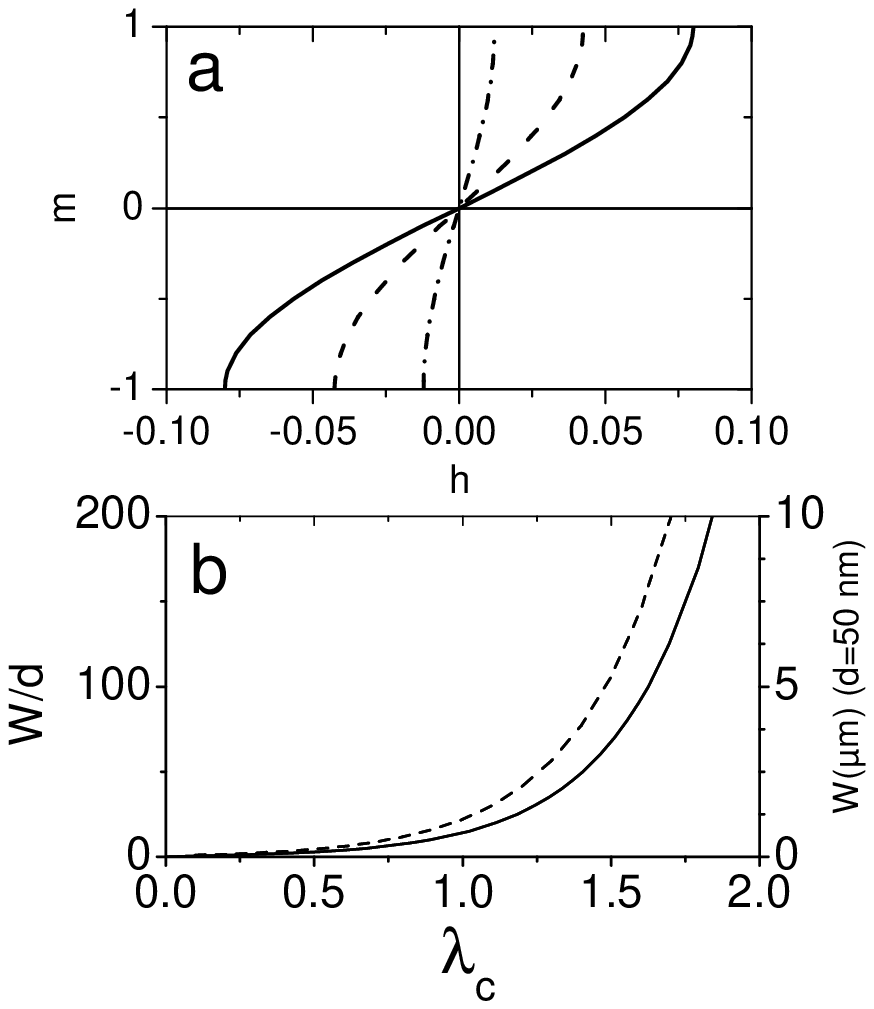}
\includegraphics[width=0.36\textwidth]{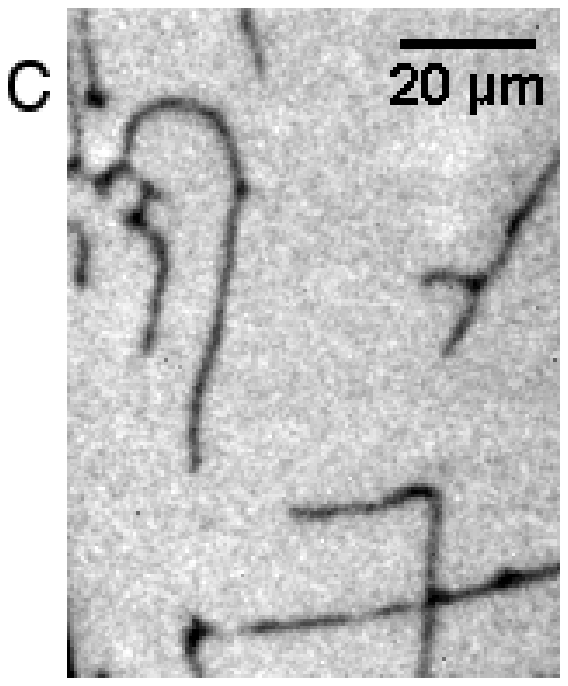}
		\label{fig:lambdaC-stripewidth}
\end{figure}

Since $Q\gg1$ domain theory for FM films with strong uniaxial anisotropy can be applied.\cite{hubert} In this framework the free energy of a stripe array $F_{str}$ is the sum of the DW energy, the Zeeman energy and intra- and inter-domain magnetic interaction terms. $F_{str}$ is a function of two variables: the reduced average magnetization along the direction of the applied field $m=\left\langle M\right\rangle/M_s$ and the stripe period $P$. Minimization of $F_{str}$ provides the dependence of $m$ and $P$ on $\lambda_c$ and on the reduced applied field $h=H/M_s$. $m(h)$ curves are shown in Fig.~\ref{fig:lambdaC-stripewidth}(a) for several values of $\lambda_c$. The stripe width of the minority phase ($M$ opposite to $H$) decreases as $h$ increases from zero up to the saturation field $h_s=h_{(m\rightarrow1)}$ (Fig.~\ref{fig:lambdaC-stripewidth}(b)). It is worth noting that this width keeps a finite value $W_s$ when $m\rightarrow1$ due to intra-domain magnetic interaction. The dependence of $W_s$ on $\lambda_c$ is obtained by solving the equation
$2\pi \lambda_c=w_s^2 \ln(1+w_s^{-2})+\ln(1+w_s^2)$,
where $w_{s}=W_s/d$. It is shown in Fig.~\ref{fig:lambdaC-stripewidth}(b). Let us note the quasi-exponential increase of the stripe width with $\lambda_c$, which means that domains cannot exist for $\lambda_c\gg 1$.

\begin{figure}[htbp]
	\begin{center}
	\includegraphics[width=0.9\linewidth]{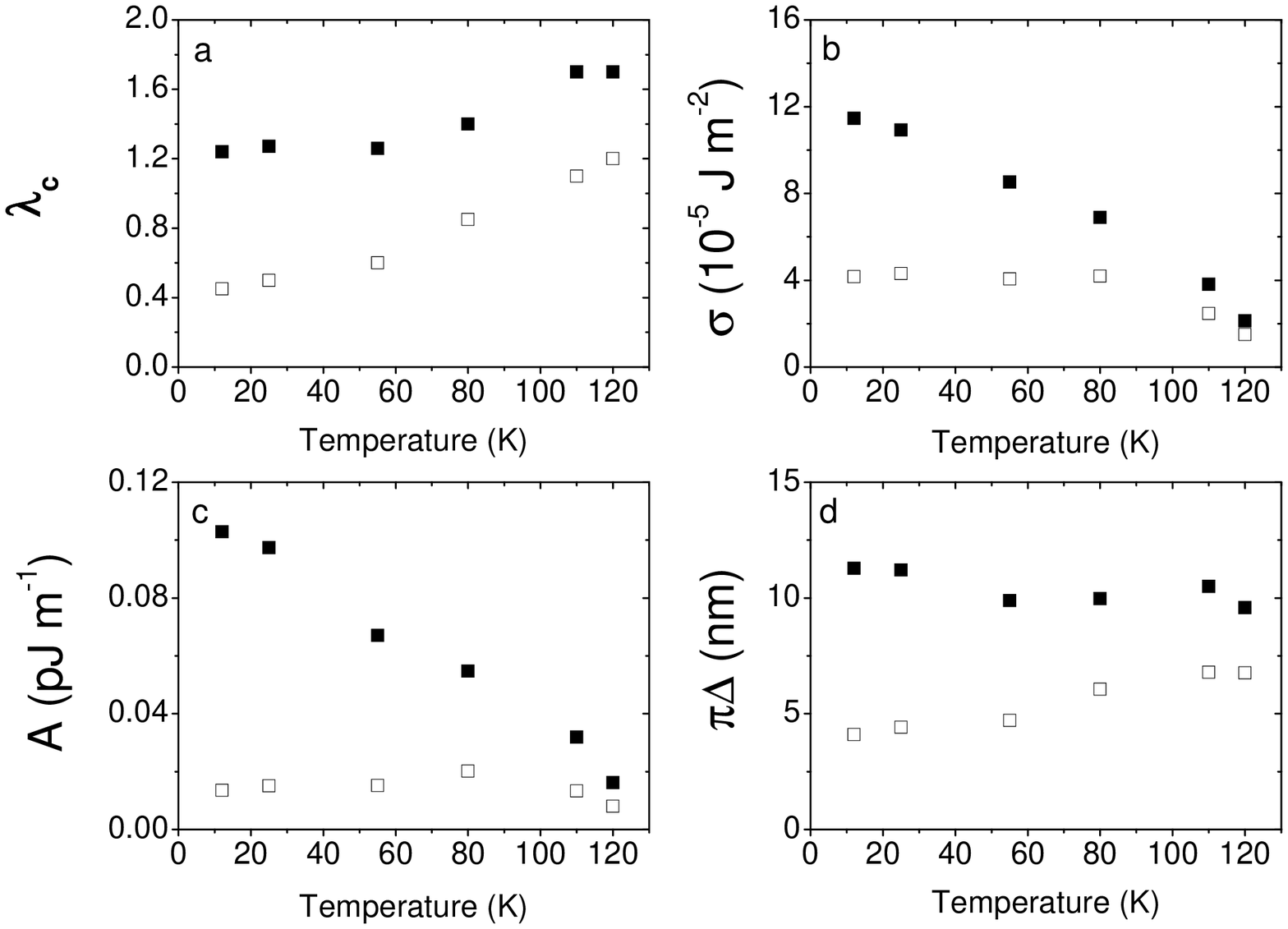}
	\end{center}
	\label{fig:graph-recap}
\end{figure}

In (Ga,Mn)As with perpendicular easy axis, DW pinning is strong enough with respect to inter-domain magnetic interaction to hinder the long-range ordering of magnetic domains.\cite{dourlatJAP} However, lamellar domains with a narrow distribution of their width are observed close to magnetization saturation, as shown in Fig.~\ref{fig:lambdaC-stripewidth}(c). This indicates that DW pinning is not sufficiently strong to prevent the formation of lamellar domains resulting from intra-domain magnetic interaction.
An upper boundary for the parameter $\lambda_c$ is obtained by comparing the prediction for the equilibrium domain width $W_s$ with the average width $W$ of these lamellar domains. At $T=80$~K, $W$ is found equal to 1.7$\pm0.3$~$\mu$m, which corresponds to $\lambda_c$=1.3. Since DW pinning impedes the reversible evolution of the domain width with the field, the width of lamellar domains should be larger than the predicted equilibrium stripe width. Moreover, owing to the limited spatial resolution ($\approx1\mu$m), the measured width is probably larger than the actual one. Finally, since $W_s$ is an increasing function of $\lambda_c$, the value of $\lambda_c$ deduced experimentally corresponds to an upper boundary for $\lambda_c$. The same analysis was repeated at different temperatures. The results are reported in Fig.\ref{fig:graph-recap}(a). A small variation of $\lambda_c^{max}$ (1.24 to 1.7) is found as the temperature $T$ varies from 12 to 120~K.

\begin{figure}
	\begin{center}
		\includegraphics[width=0.9\linewidth]{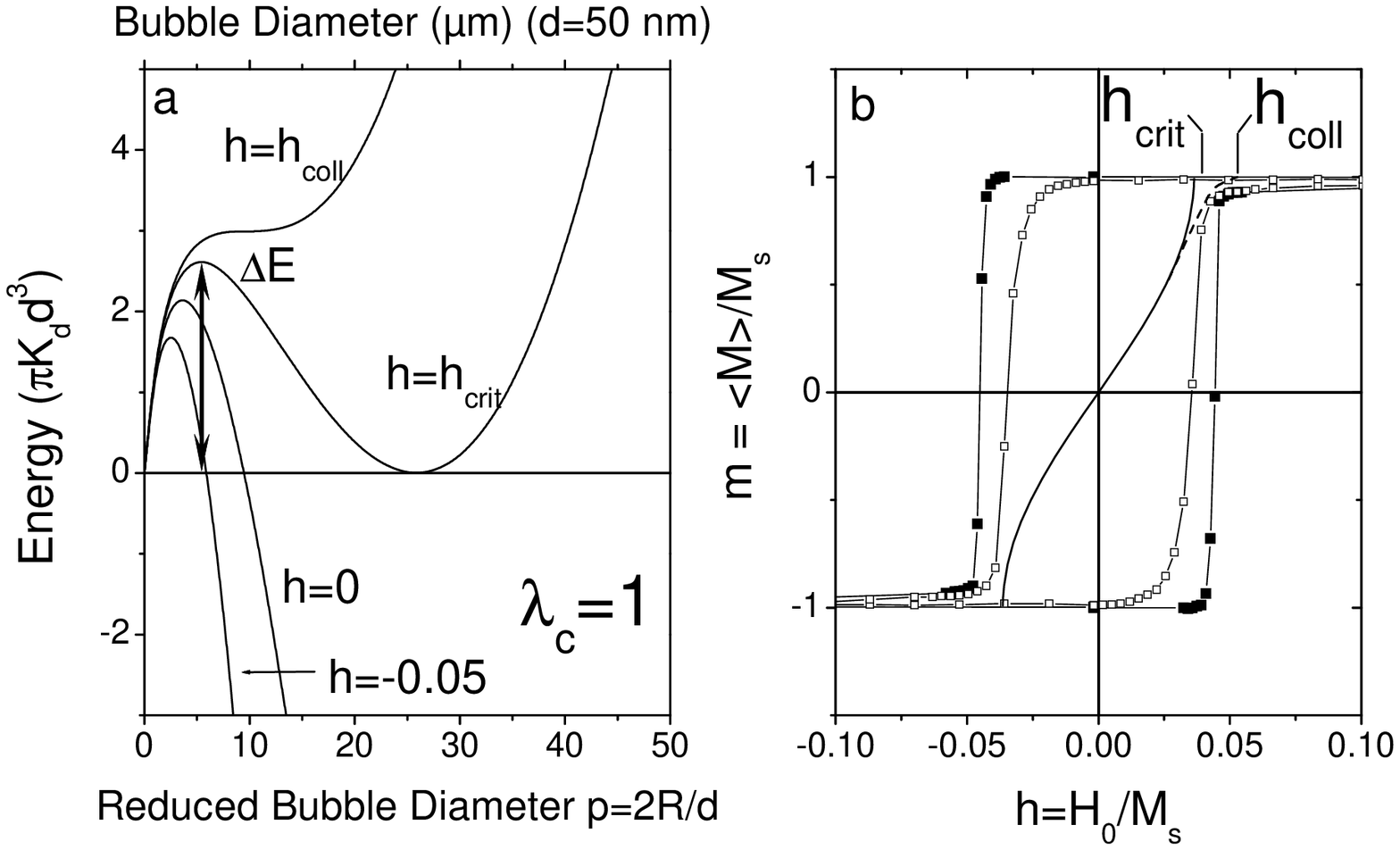}
		\end{center}
	\label{fig:graph-energy}
\end{figure}

A lower boundary for $\lambda_c$ is determined by comparing the theoretical predictions for the magnetization curve $m(h)$ with the hysteresis cycles obtained experimentally. Because of hysteresis, the $m(H)$ curve predicted for thermodynamical equilibrium and without pinning should lie \textit{inside} the experimental hysteresis cycle. A major hysteresis cycle at $T$=80~K is shown in Fig.~\ref{fig:graph-energy}(b) (black squares). The applied field is swept up to 60~mT, \textit{i.e.}, beyond the saturation field ($\approx$~12 mT), which is much larger than the coercive field $\mu_0H_c=1.6$~mT. When the field is swept down, the onset of magnetization reversal and therefore the width of the cycle is determined by defect-assisted nucleation (see below). In order to obtain a narrower cycle, we recorded a minor hysteresis loop (empty squares). It is obtained by sweeping the field up to $\mu_0H_m=6$~mT, \textit{i.e.}, below the saturation field, in order to leave a few reversed domains in the sample. Hence, upon sweeping the field down, the nucleation stage is avoided. The shape of the hysteresis loop is determined by the dependence of the DW velocity on the applied field. Care is taken to use a low field ramp rate in order to allow for DW propagation at low velocity in the vicinity of the coercive field. The shape and width of the minor hysteresis loop does not depend on $H_m$ provided it is larger than the coercive field. The theoretical $m(h)$ curves for stripes were calculated for a set of $\lambda_c$ values (Fig.~\ref{fig:lambdaC-stripewidth}(a)). The saturation field increases for decreasing $\lambda_c$. Therefore, by selecting the $m(h)$ curve which is tangent to the hysteresis cycle close to saturation, one can obtain a lower boundary for $\lambda_c$. This analysis can be further refined by considering that the saturation field calculated for stripes is lower than the theoretical collapse field, i.e., the field at which domains of the minority phase disappear. As $m\rightarrow1$, lamellar domains keep a finite width but their length decreases until they reach the bubble circular shape and finally collapse.\cite{gourdonPRL2006} The bubble collapse field $h_{coll}$ can be calculated from the free energy of an isolated bubble with reverse magnetization\cite{hubert,gourdonPRL2006}  
\begin{equation}
F_{sb}=\pi K_d d^3 \left[2 \lambda_c p +(h-1)p^2+ p^2 f(p)\right]\;,
\label{single-bubble}
\end{equation}
where $p=2R/d$ is the reduced diameter of the bubble and $f(p)=1+4p\left\{1-\left[(2k^2-1)E(k)+(1-k^2)K(k)\right]k^{-3}\right\}/3\pi$
with $k^2=p^2/(1+p^2)$. $E$ and $K$ are complete elliptic integrals of the first and the second kind, respectively. $F_{sb}$ varies non-monotonously  with $p$ as shown in Fig.~\ref{fig:graph-energy}(a). Above the critical field $h_{crit}$ the system remains is a metastable state until the collapse field is reached. $h_{coll}$ is the field at which $\partial F_{sb}/\partial p=\partial^2 F_{sb}/\partial p^2=0$. It increases with increasing $\lambda_c$.

Figure~\ref{fig:graph-energy}(b) shows the the critical and collapse fields. The dashed curve  shows the expected magnetization path from the stripe array curve to the bubble collapse field for the metastable domain pattern. The lower boundary for $\lambda_c$ is taken as the value that makes this curve tangent to the minor hysteresis cycle close to saturation.
At $T=80$~K, $\lambda_c^{min}$=0.85 is obtained. This procedure was repeated at different temperatures.

Finally we obtain the micromagnetic parameters. They are displayed in Figure~\ref{fig:graph-recap}. The specific wall energy $\sigma$, the DW width $\pi\Delta=\pi\sqrt{A/K_u}$, and the exchange spin stiffness constant $A$ are derived from the $\lambda_c$ values using the temperature dependence of $M_s$ and $K_u$. $\sigma$ is found in the range 1.5-11.5~10$^{-5}$~J~m$^{-2}$, which is about 2 orders of magnitude smaller than for iron or cobalt films.\cite{huoJAP,chenJAP} This explains that propagating DWs easily skirt around pinning defects.\cite{thevenard,dourlatJAP} The DW width is found in the range 4-11 nm. This is larger than the mean distance between Mn ions (0.9 nm for 7\% Mn) and smaller than the domain width, which validates the use of domain theory. The ratio of the DW parameter to the sample thickness $\Delta/d$ is larger than 0.025. Together with the large $Q$ value, this ensures that the specific DW energy is very close to $4\sqrt{A K_u}$, as assumed in our analysis.~\cite{hubertJAP75}

From the micromagnetic parameters one can estimate the Walker breakdown velocity $V_W$ for field-driven DW propagation in the flow regime.\cite{hubert} Using $V_W=\gamma\sqrt{2\mu_0AQ}(\sqrt{1+1/Q}-1)\approx\gamma\mu_0M_s\Delta/2$ with $\gamma$ the gyromagnetic ratio, one finds  $V_W\approx5.5$~m~s$^{-1}$ for the lower $\Delta$-values and  15 to 9.6~m~s$^{-1}$ for the largest $\Delta$-values for $T$ between 12 and 80~K. These last $V_W$ values are consistent with the largest velocities measured in the depinning regime for the same sample.\cite{dourlat-spintech} This indicates that the values of $\lambda_c$, $\sigma$, $A$, and $\Delta$ are very likely the largest ones, \textit{i.e.}, those determined from the lamellar domain width.

From the determination of $\lambda_c$, one also gains some insight into the magnetization reversal process. We obtain the height of the intrinsic nucleation barrier. As seen in Fig.~\ref{fig:graph-energy}(a), below the critical field the energy of the system with a reverse bubble domain is lower than the energy of the saturated state. However the system can stay in the saturated state since there is an energy barrier $\Delta E$ for nucleation. Taking 0.85$<\lambda_c<$1.3 at $T=80$~K yields 280$<\Delta E/KT<$1310 at $h=h_{crit}$. The barrier height only weakly decreases with decreasing field. At the onset of magnetization reversal ($h\approx-0.05$) one still finds large values for $\Delta E$, namely 194$<\Delta E/KT<$660. This means that the system is highly metastable and that nucleation proceeds through local defects that strongly decrease the barrier height. This conclusion also holds at $T=120$~K, \textit{i.e.}, close to $T_C$. 

The spin stiffness constant is smaller by one order of magnitude than the value determined by Goennenwein \textit{et al.} from spin wave resonances in FMR spectra for (Ga,Mn)As with in-plane magnetization.\cite{goennenwein} Let us note that a linear gradient of the magnetic properties had to be assumed in [\onlinecite{goennenwein}] in order to fit the results with theoretical spin wawe dispersion curves. Theoretical predictions for $A$ were made in Ref.~[\onlinecite{konig}] for various carrier densities and for two values of the $J_{pd}$ exchange constant, taking a Mn concentration of 4.5\%. For our sample we obtain an effective Mn concentration $x_{eff}=3.8$\% (we neglect the hole magnetization and assume $M_s(T\rightarrow0)=x_{eff}N_0 5\mu_B$, with $N_0$ the density of cations site). With this Mn concentration, our results for the spin stiffness are consistent with the extrapolated theoretical estimations of [\onlinecite{konig}] provided that a low value of the $J_{pd}$ exchange constant is used ($J_{pd}\leq50$~meV~nm$^3$).


To summarize, we have applied domain theory to the case of a metastable system in order to obtain the ratio of DW specific energy and magnetic energy. Combining the study of domain structure, magnetization and anisotropy constants in a (Ga,Mn)As thin film with perpendicular magnetization, we have determined the micromagnetic parameters: the specific DW energy (1.5-11.5~10$^{-5}$~J~m$^{-2}$), the DW width (4-11.3~nm) and the spin stiffness constant(0.01-0.1~pJ~m$^{-1}$). From these results we estimate the nucleation barrier and the Walker velocity for DW propagation.


\newpage
\begin{center}
FIGURE CAPTIONS
\end{center}

FIG.1 (a) Temperature dependence of the anisotropy contants $K_{2\bot}$ (black squares), $K_{4\bot}$ (black circles), $K_{2\|}$ (open squares), $K_{4\|}$ (open circles). $K_u=K_{2\bot}+K_{4\bot}$ (black diamonds, dashed line). (b) Temperature dependence of the magnetization (black squares, left scale) and of the $Q$ parameter (black circles, right scale).\\

FIG. 2 (a) Dependence of the reduced average magnetization $m=\left\langle M\right\rangle/M_s$ on $h=H/M_s$ for a stripe array with $\lambda_c=0.6$ (full curve), $0.8$ (dashed), $1.2$ (dash-dotted). (b) Stripe width as a function of the parameter $\lambda_c$ for stripes of the minority phase near saturation (solid curve) and stripes at $m=0$, $H=0$ (dashed curve). (c) Lamellar domain structure in (Ga,Mn)As near the saturation field at T=80~K.\\

FIG. 3 Temperature dependence of the upper (black squares) and lower (empty squares) boundaries of : (a) the parameter $\lambda_c$, (b) the specific wall energy $\sigma$, (c) the exchange spin stiffness $A$,  and (d) the DW width $\pi\Delta$. They are obtained from the lamellar domain width (upper values) and hysteresis cycle (lower values).\\

FIG.4 (a) Energy of an isolated bubble with reverse magnetization for several magnetic fields ($\lambda_c$=1). (b) Major (black squares) and minor (empty squares) hysteresis cycles obtained from Kerr microscopy images at $T$=80~K. The full curve is the $m(h)$ curve for a stripe array with $\lambda_c$=0.85. The bubble critical and collapse fields for $\lambda_c$=0.85 are indicated as $h_{crit}$ and $h_{coll}$, respectively. The dashed curve links the $m(h)$ curve for stripes to the bubble collapse point ($h_{coll}$, $m=1$). It shows the expected magnetization path resulting from the existence of the collapse barrier.

\newpage
\begin{center}
FIGURE 1

\end{center}

\newpage
\begin{center}
FIGURE 2

\end{center}

\newpage
\begin{center}
FIGURE 3

\end{center}

\newpage
\begin{center}
FIGURE 4

\end{center}

\end{document}